\documentclass[12pt]{iopart}

\usepackage{graphicx}

\begin{document}

\title{Adaptive financial networks with static and dynamic thresholds}

\author{Tian Qiu$^1$, Bo Zheng$^2$ and Guang Chen$^1$}

\address{$^1$School of Information Engineering, Nanchang
Hangkong University, Nanchang 330063, P.R. China \\$^2$Zhejiang
Institute of Modern Physics, Zhejiang University, Hangzhou 310027,
P.R. China} \ead{\mailto{tianqiu.edu@gmail.com},
\mailto{zheng@zimp.zju.edu.cn} and \mailto{cgcgchen@gmail.com}}
\begin{abstract}
Based on the daily data of American and Chinese stock markets, the
dynamic behavior of a financial network with static and dynamic
thresholds is investigated. Compared with the static threshold, the
dynamic threshold suppresses the large fluctuation induced by the
cross-correlation of individual stock prices, and leads to a stable
topological structure in the dynamic evolution. Long-range
time-correlations are revealed for the average clustering
coefficient, average degree and cross-correlation of degrees. The
dynamic network shows a two-peak behavior in the degree
distribution.

\end{abstract}

\maketitle

\section{Introduction}

A financial market is a complex system composed of many interacting
units, which exhibits various collective behaviors
\cite{man95,gop99,lux99,gia01,ren06,qiu06,she09a}. How to extract
its structure information has attracted much attention of
physicists, and remains challenging. For example, the hierarchical
structure of financial markets has been investigated with the
minimal spanning tree and its variants \cite
{man99,bon03,mic03,tum05,tum07}. With the random matrix theory,
business sectors and unusual sectors can be clarified, and topology
communities are also revealed
\cite{ple02,she09,cor05,pan07,uts04,gar08}.

The complexity theory provides a powerful tool to understand complex
networks. In recent years much progress has been achieved on the
fields from biology to sociology \cite{alb99,alb02,cal00,sat01}.
Different models and theoretical approaches have been developed to
investigate the formation of network topologies, the functions
emerging from networks, the combined mechanism of topologies and
functions, etc \cite{bar99,wat98,may01,zhu08,gro08}. Most efforts of
these schemes focus on the collective properties of the system in a
steady state. However, complex systems such as the financial markets
are essentially nonstationary in time. For each time step, the
system shows a particular topology, induced by the
cross-correlations of individual stock prices. The topology dynamics
is important for the full understanding of the network structure.
However, it is rarely touched or elaborated in detail, to the best
of our knowledge.

So far, static topological properties of financial markets from the
view of complex networks have been widely investigated
\cite{onn04,hua09,kim02,son09,yan08,kim07}. An ordinary way to
construct a financial network is to take individual stocks as nodes,
and set a threshold to create edges. If the cross-correlation
between two stocks is larger (smaller) than the threshold, then link
(cut) an edge. A number of contributions have adopted an
artificially given threshold for static cross-correlations. However,
it does not extract the dynamic characteristics of the stock markets
\cite{hua09,yan08,kim07}. For example, an outstanding company with a
great number of correlated companies is driven bankrupt by an
extreme event, and all its linkages are then cut down. This may lead
to a temporal variation of the topological structure of the market.
Therefore, a real financial network should be dynamic, to capture
the dynamic evolution of the topological structure. On the other
hand, as is well known, the volatilities of the stock prices exhibit
a fat-tailed probability distribution. Large fluctuations of the
volatilities may greatly influence the network structure and network
stability, and therefore should be taken into account in
understanding the topology dynamics.

In this article, we examine a dynamic financial network based on the
American and Chinese stock markets. For a comparative study, both
static and dynamic thresholds are respectively adopted in the
network construction. Our purpose is to study the statistical
properties of the dynamic network, and more importantly, to
investigate the temporal correlations of the topology time series,
such as the time series of the average clustering coefficient, the
average degree and the cross-correlation of degrees, by applying the
detrended fluctuation analysis (DFA). Special attention is put on
the dynamic effect of the thresholds on the network structure and
network stability.

In section II, we present the data analyzed and construct the
dynamic network with static and dynamic thresholds. In section III,
we investigate the time-correlations of the average clustering
coe¡Àcient, the average degree and the cross-correlation of degrees,
and examine the degree distribution of the network. Finally, the
conclusion comes in section IV.

\section{Dynamic financial network}

For a comprehensive understanding, we analyze two different stock
markets, the New York Stock Exchange (NYSE) and the Chinese Stock
Market (CSM), representing the mature and the emerging markets,
respectively. For both markets, we investigate the daily data of 259
individual stocks, with 2981 data points from the year 1997 to 2008
for the NYSE, and 2633 data points from the year 1997 to 2007 for
the CSM.

We define the price return
\begin{equation}
R_{i}(t,\Delta{t})=\ln{P_{i}(t+\Delta t)}-\ln{P_{i}(t)}, \label{e10}
\end{equation}
where $P_{i}(t)$ is the price of stock $i$ at time $t$, and the time
interval is set to $\Delta{t}=1$ day in this article. For comparison
of different stocks, we normalize the price return to
\begin{equation}
r_{i}(t) = \frac {R_{i}-<R_{i}>} {\sigma_{i}}, \label{e20}
\end{equation}
where $\sigma_{i} = \sqrt{<R_{i}^{2}>-<R_{i}>^{2}}$ is the standard
deviation of $R_{i}$, and $<\ldots>$ is the time average over $t$.
We define an instantaneous equal-time cross-correlation between two
stocks by
\begin{equation}
G_{ij}(t)=r_{i}(t)r_{j}(t), \label{e30}
\end{equation}

\begin{figure}[htb]
\centering
\includegraphics[width=8.5cm]{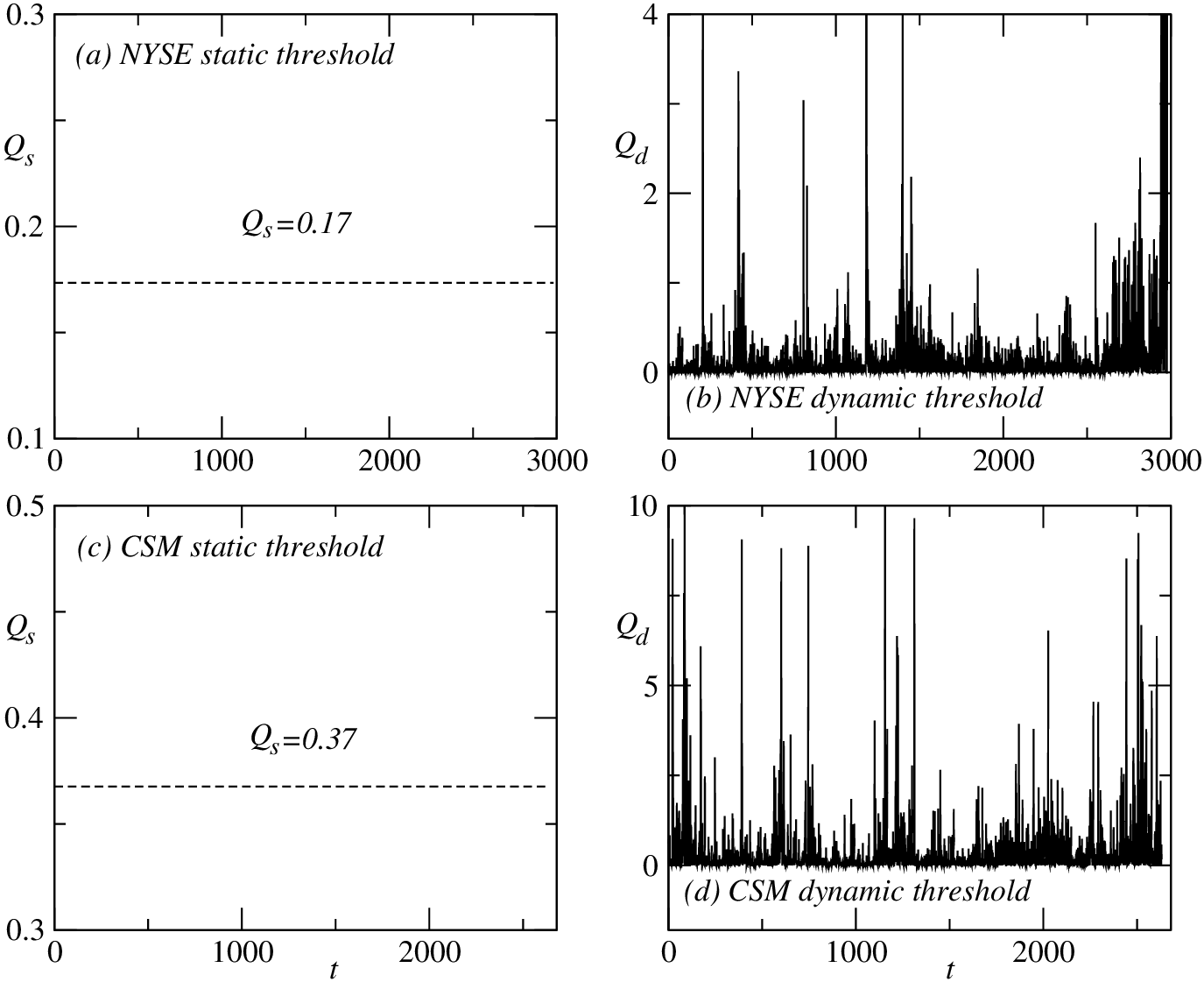}
\caption{\label{Fig:1} $Q_{s}$ and $Q_{d}(t)$ for the NYSE and CSM
are shown in (a), (b), (c) and (d).}
\end{figure}

Our financial network is constructed in the following way: take
individual stocks as nodes and set a threshold $\zeta$ to create
edges. At each time step, if the cross-correlation
$G_{ij}(t)>\zeta$, then add an edge between stocks $i$ and $j$;
otherwise, cut the edge. Due to the dynamic evolution of
$G_{ij}(t)$, the connections between stocks vary with time,
regardless of the static or the dynamic threshold. This results in a
dynamic topology of the financial network. To understand the
robustness and stability of the network structure, introducing a
proper threshold is very important. We first consider a static
threshold $\zeta \sim Q_s$,
\begin{equation}
Q_{s}=\frac{2}{N(N-1)T}\sum _{i=1}^{N} \sum _{j=i+1}^{N} \sum
_{t=1}^{T} G_{ij}(t), \label{e40}
\end{equation}
where $T$ is the total time interval, and $N$ is the total stock
number. Since $Q_{s}$ is the average cross-correlation over the
total time and all stocks, the static threshold is reasonable in the
sense of providing a uniform standard for cross-correlations of all
stocks. As shown in figure \ref{Fig:1}(a) and (c), $Q_{s}$ takes
rather small values, 0.17 for the NYSE and 0.37 for the CSM.
However, the cross-correlation $G_{ij}(t)$ is defined by the
individual price returns, and fluctuate according to the price
dynamics. The static threshold may suffer from the large fluctuation
of the cross-correlation $G_{ij}(t)$. Hence, we introduce a dynamic
threshold $\zeta \sim Q_d(t)$,
\begin{equation}
Q_{d}(t)=\frac{2}{N(N-1)}\sum _{i=1}^{N} \sum _{j=i+1}^{N}
G_{ij}(t), \label{e50}
\end{equation}
$Q_{d}(t)$ takes only the average over all stocks in a single time
step, and therefore fluctuates synchronously with the
cross-correlations $G_{ij}(t)$. As shown in figure \ref{Fig:1}(b)
and (d), large values of $Q_{d}(t)$ are observed for both the NYSE
and the CSM, in addition to the small values for most time steps.
Therefore, the dynamic threshold may suppress the large fluctuation
induced by $G_{ij}(t)$, and accordingly creates a stable network
structure. In this paper, we consider the static threshold from
$\zeta=0.25Q_{s}$ to $4Q_{s}$, and the dynamic threshold from
$\zeta=0.25Q_{d}$ to $6Q_{d}$.

\begin{figure}[htb]
\centering
\includegraphics[width=8.5cm]{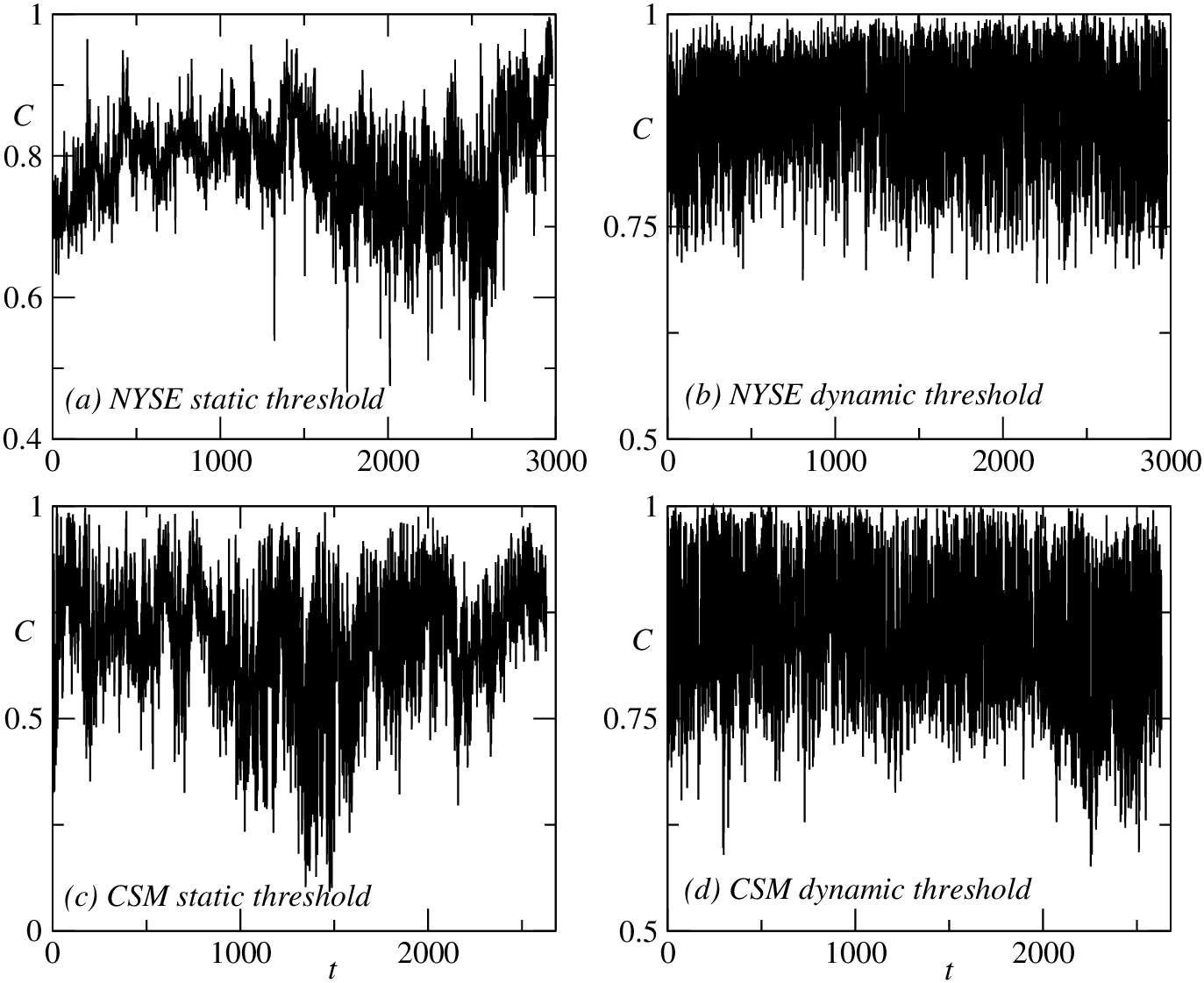}
\caption{\label{Fig:2} The average clustering coefficients $C(t)$
corresponding to $Q_{s}$ and $Q_{d}(t)$ for the NYSE and CSM are
displayed in (a), (b), (c) and (d). The time-averaging values of
$C(t)$ are $0.78$, $0.88$, $0.68$ and $0.85$ respectively.}
\end{figure}

In figure \ref{Fig:2}, the average clustering coefficient $C(t)$ is
displayed for $\zeta=Q_{s}$ and $\zeta=Q_{d}(t)$. The average
clustering coefficient $C(t)$ is defined as
\begin{equation}
C(t)=\frac{1}{N}\sum_{i=1}^{N}c_{i}(t), \label{e60}
\end{equation}
where $c_{i}(t)$ is the clustering coefficient of node $i$
\cite{wat98}, denoting the ratio of the triangle-connection number
of node $i$ to the maximum possible triangle-connection number of
the node. It is observed in figure \ref{Fig:2} that the average
clustering coefficient $C(t)$ of $\zeta=Q_{d}(t)$ is apparently more
stable than that of $\zeta=Q_{s}$ for both the NYSE and the CSM. The
common characteristic shared by $\zeta=Q_{s}$ and $\zeta=Q_{d}(t)$
is that their average clustering coefficients are large. Especially
for the dynamic threshold $\zeta=Q_{d}(t)$, the average clustering
coefficient $C(t)$ keeps its values around 0.88 and 0.85
respectively for the NYSE and the CSM at a highly clustering level,
implying a close relationship of stocks in the financial markets.

\section{Topology dynamics}

Topological properties of a complex network are usually described by
the average clustering coefficient, the average degree and the
cross-correlation of degrees. We investigate the topology dynamics
by computing the time-correlations. The autocorrelation function is
widely adopted to measure the time-correlation. However, it shows
large fluctuations for nonstationary time series. Therefore, we
apply the DFA method \cite {pen94,pen95}.

For a time series $A(t')$, we eliminate the average trend from the
time series by introducing $B(t')=\sum\limits_{t''=1}^{t'}
[A(t'')-A_{ave}]$, where $A_{ave}$ is the average of $A(t')$ in the
total time interval $[1, T]$. Uniformly dividing $[1, T]$ into
windows of size $t$, and fitting $B(t')$ to a linear function
$B_{t}(t')$ in each window, we define the DFA function as
\begin{equation}
F(t)=\sqrt{\frac{1}{T}\sum\limits_{t'=1}^{T}{[B(t')-B_t(t')]}^2},
\label{e70}
\end{equation}
In general, $F(t)$ will obey a power-law scaling behavior $F(t)\sim
t^{\theta}$. The exponent $\theta>1.0$, $0.5<\theta<1.0$ and
$0<\theta<0.5$ indicate an unstable, long-range correlating and
anti-correlating time series, respectively. $\theta=0.5$ corresponds
to the Gaussian white noise, while $\theta=1.0$ represents the $1/f$
noise.

\subsection{Average clustering coefficient}

\begin{figure}[htb]
\centering
\includegraphics[width=8.5cm]{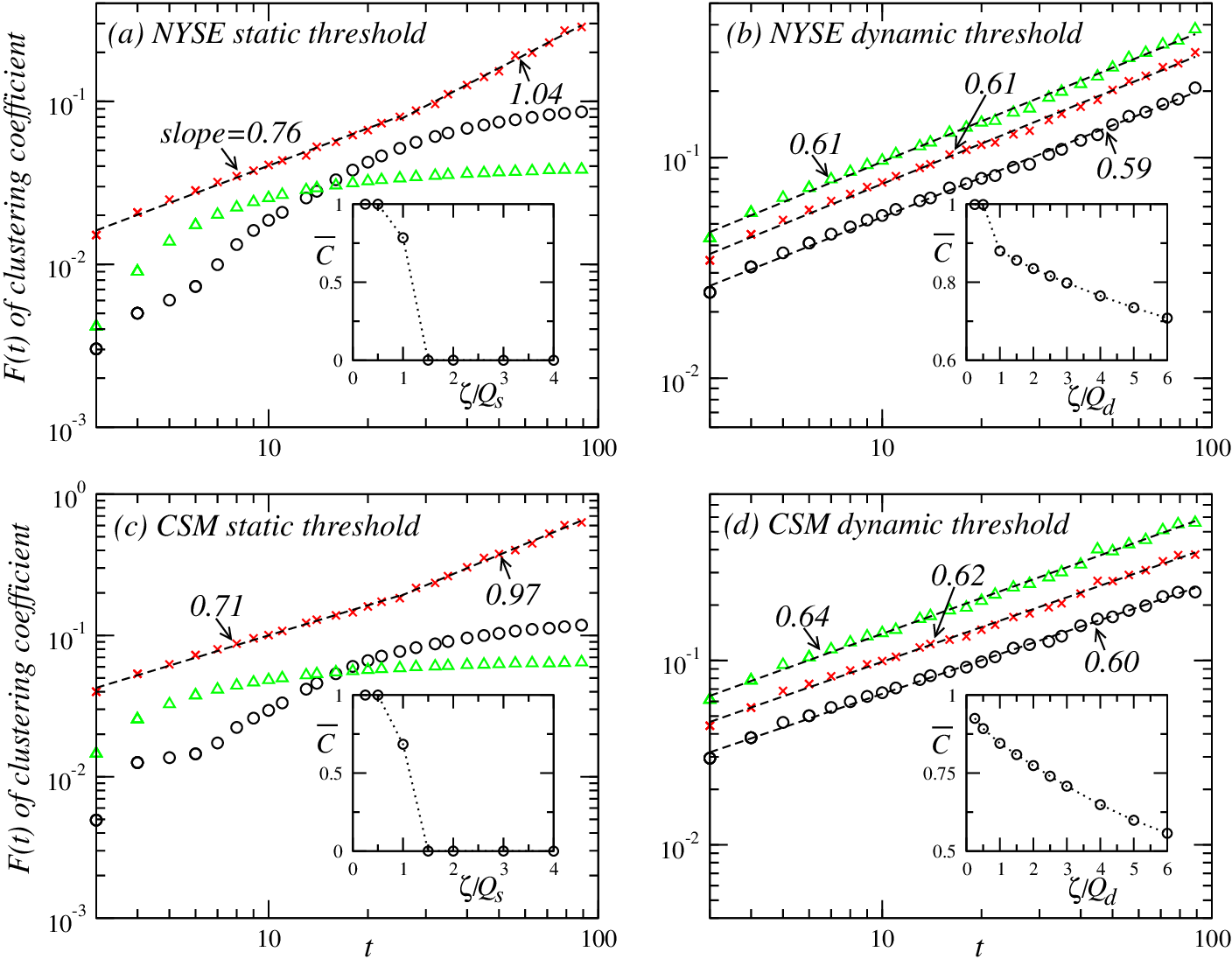}
\caption{\label{Fig:3} The DFA function of the average clustering
coefficient $C(t)$ is plotted on a log-log scale. The inner panel
shows the time-averaging clustering coefficient for different
threshold values. Dashed lines are the power-law fits. (a) and (c)
are of the static thresholds for the NYSE and the CSM. Circles,
crosses and triangles are for $\zeta=0.5Q_{s}$, $Q_{s}$ and
2$Q_{s}$, respectively. (b) and (d) are of the dynamic thresholds
for the NYSE and the CSM. Circles, crosses and triangles are for
$\zeta=Q_{d}(t)$, 2$Q_{d}(t)$ and 3$Q_{d}(t)$, respectively.}
\end{figure}

The DFA function of the average clustering coefficient $C(t)$ is
shown in figure \ref{Fig:3}. For the static threshold $\zeta=Q_s$, a
two-stage scaling behavior is observed with a crossover phenomenon
in between. For the NYSE, the exponent $\theta$ takes the values
$\theta=0.76$ for $t<t_{c}$ and $\theta=1.04$ for $t>t_{c}$. For the
CSM, $\theta=0.71$ for $t<t_{c}$ and $\theta=0.97$ for $t>t_{c}$.
The crossover time $t_{c}\sim 25$ days. This result indicates that
the average clustering coefficients are temporally correlated for
$t<t_{c}$, then transit to the $1/f$ noise for $t>t_{c}$. In
contrast to that of $\zeta=Q_{s}$, the DFA function of
$\zeta=Q_{d}(t)$ shows a clean power-law behavior, with the exponent
$\theta=0.59$ for the NYSE and $0.60$ for the CSM, indicating the
long-range time correlation.

To investigate the robustness of the above results, and the
stability of the network structure during the dynamic evolution, we
may adjust the level of the thresholds. For the static threshold, we
consider two alternatives $\zeta=0.5 Q_{s}$ and $\zeta=2Q_{s}$. As
shown in figure \ref{Fig:3}(a) and (c), one does not find any
power-law behavior. In other words, the network structure is rather
sensitive to the specific value of the static threshold. For the
dynamic threshold, we observe that the DFA function of $C(t)$
remains qualitatively the same for $\zeta \ge Q_d(t)$. In figure
\ref{Fig:3} (b) and (d), for example, the results of $\zeta=2Q_{d}$
and $3Q_{d}$ are displayed. A clean power-law behavior is detected
with the exponent $\theta=0.61$ and $0.61$ for the NYSE, and
$\theta=0.62$ and $0.64$ for the CSM. Higher dynamic thresholds
yield stronger time correlations (i.e., larger values of $\theta$).
This implies that the edges created by large $G_{ij}(t)$ are rather
stable.

To further understand the topological pattern, we calculate the
time-averaging clustering coefficient $\overline{C}$ for different
thresholds,
\begin{equation}
\overline{C}=\frac{1}{T}\sum_{t=1}^{T}C(t), \label{e80}
\end{equation}
As shown in the inner panel of figure \ref{Fig:3}(a) and (c),
$\overline{C}$ of the static threshold $\zeta \le 0.5Q_{s}$ is close
to 1, but as the threshold increases to 1.5$Q_{s}$, $\overline{C}$
sharply drops to 0. As shown in the inner panel of figure
\ref{Fig:3}(b) and (d), the time-averaging clustering coefficient
$\overline{C}$ shows a mild decay for the dynamic threshold, and
presents still rather high clustering even for the threshold
$\zeta=6Q_{d}(t)$ for both the NYSE and the CSM.

\subsection{Average degree}

\begin{figure}[htb]
\centering
\includegraphics[width=8.5cm]{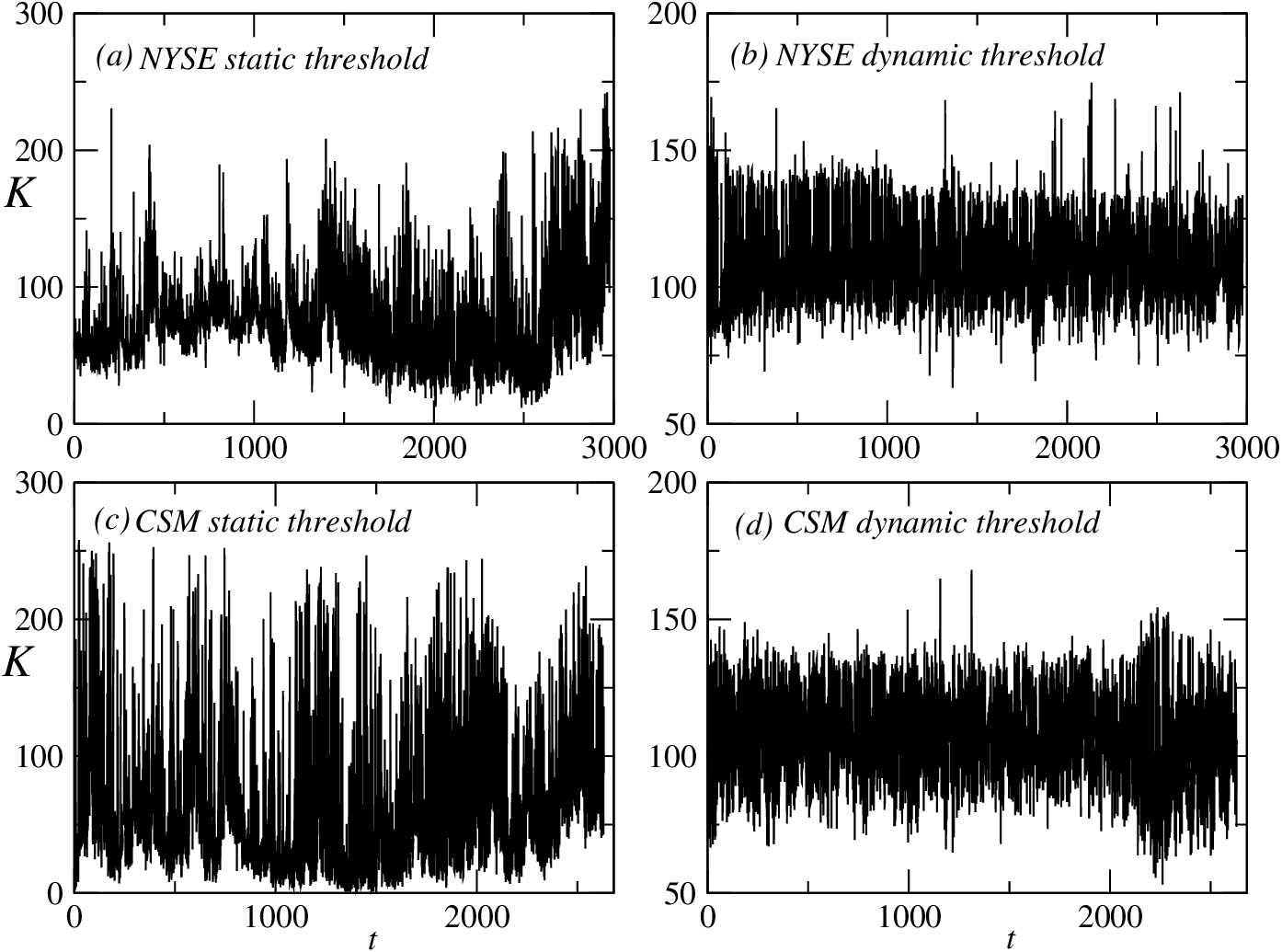}
\caption{\label{Fig:4}The average degree $K(t)$ is displayed. (a)
and (c) are of the static threshold $\zeta=Q_{s}$ for the NYSE and
the CSM. (b) and (d) are of the dynamic threshold $\zeta=Q_{d}(t)$
for the NYSE and the CSM.}
\end{figure}

The average degree $K(t)$ of the network is defined as
\begin{equation}
K(t)=\frac{1}{N}\sum_{i=1}^{N}k_{i}(t), \label{e90}
\end{equation}
where $k_{i}(t)$ is the degree of node $i$, denoting the number of
the nodes directly connected with it. For $\zeta=Q_{s}$, $K(t)$
exhibits large fluctuations between $0$ and $N-1$, as shown in
figure \ref{Fig:4}(a) and (c). ${K(t)}=N-1$ indicates that every
node directly connects to all other nodes in the network, while
${K(t)}=0$ corresponds a set of isolated nodes. The fluctuation of
the CSM is obviously stronger, and $K(t)$ often comes rather close
to $0$ and $N-1$. For $\zeta=Q_{d}(t)$, $K(t)$ is almost bounded by
two envelope curves, as shown in figure \ref{Fig:4}(b) and (d). The
upper envelope is around $K(t) \sim 130$ to $140$, indicating that
one node is directly connected to about one half of the nodes on
average, while the lower envelope is around $K(t) \sim 70$ to $80$,
indicating that one node is directly connected to about one quarter
of the nodes on average. The upper envelope implies a high-low
symmetric distribution of the cross-correlations $G_{ij}(t)$, with
one half of the data points below the average and the other half of
the data points above the average. The upper envelope sounds
reasonable, for the larger $G_{ij}(t)$ are usually well above the
average. However, the lower envelope suggests a high-low asymmetric
distribution of the cross-correlations $G_{ij}(t)$. Why there is
such a lower envelope remains to be understood.

\begin{figure}[htb]
\centering
\includegraphics[width=8.5cm]{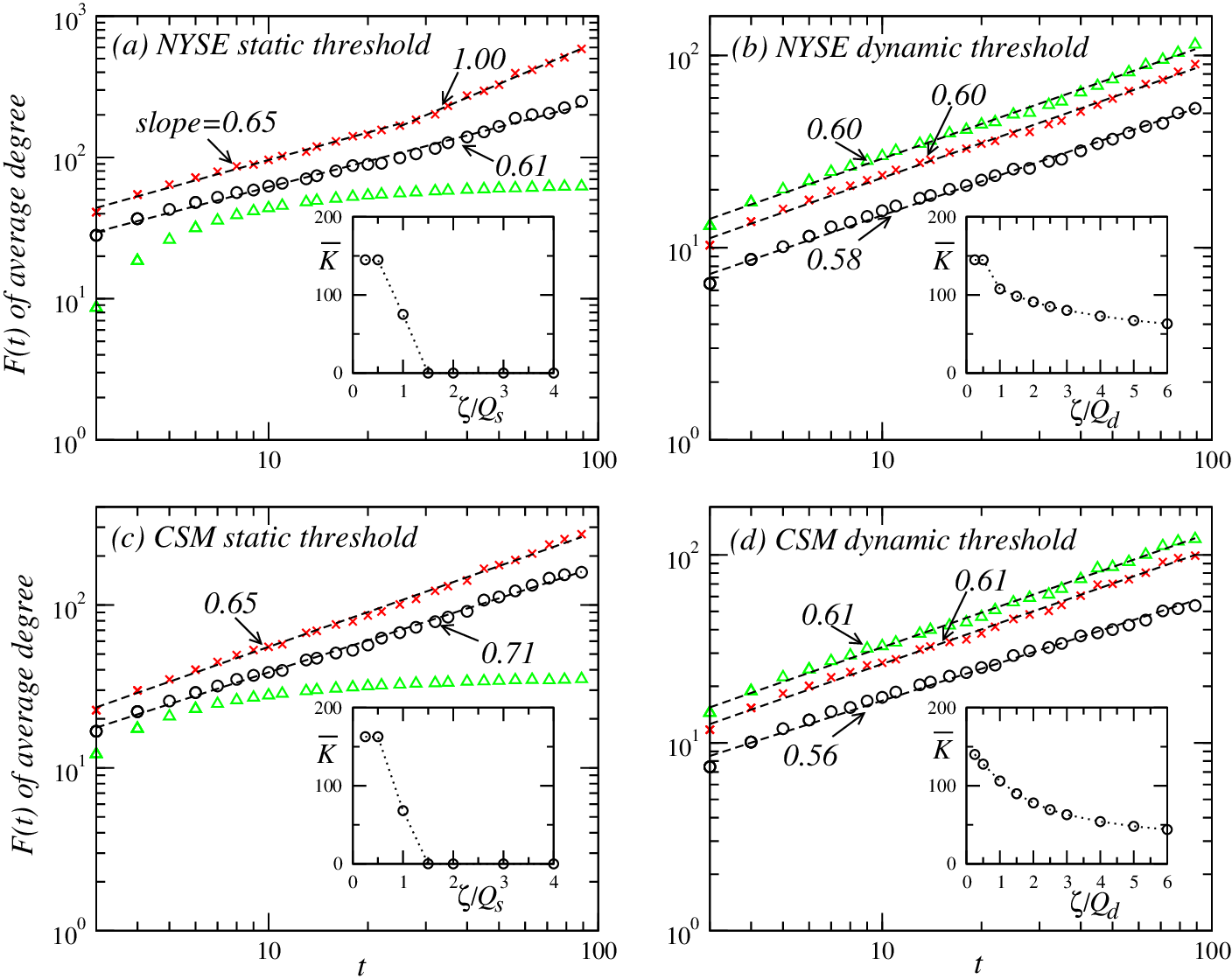}
\caption{\label{Fig:5} The DFA function of the average degree $K(t)$
is plotted on a log-log scale. Inner panels show the time-averaging
degrees for different threshold values. Dashed lines are the
power-law fits. (a) and (c) are of the static thresholds for the
NYSE and the CSM. Circles, crosses and triangles are for
$\zeta=0.5Q_{s}$, $Q_{s}$ and 2$Q_{s}$, respectively. (b) and (d)
are of the dynamic thresholds for the NYSE and the CSM. Circles,
crosses and triangles are for $\zeta=Q{_d}(t)$, 2$Q{_d}(t)$ and
3$Q_{d}(t)$, respectively.}
\end{figure}

We then compute the DFA function of the average degree $K(t)$.
Similar as that of the average clustering coefficient, unstable
behavior is detected for the static threshold, whereas stable
network structure is observed for the dynamic threshold, as shown in
figure \ref{Fig:5}. For the static threshold, the DFA function for
the NYSE shows a power-law behavior with the exponent $\theta=0.61$
for $\zeta=0.5Q_{s}$, a two-stage scaling for $\zeta=Q_{s}$, and no
power-law behavior for $\zeta=2Q_{s}$. The DFA function for the CSM
shows a power-law behavior with the exponent $\theta=0.65$ and 0.71
for $\zeta=0.5Q_{s}$ and $Q_{s}$, and no power-law behavior for
$\zeta=2Q_{s}$. These results can also be understood from the
time-averaging degree
\begin{equation}
\overline{K}=\frac{1}{T} \sum _{t=1}^{T} K(t), \label{e110}
\end{equation}
As shown in the inner panel of figure \ref{Fig:5}(a) and (c), the
network has a large number of edges for lower static thresholds,
however, the number of edges sharply falls off to nearly zero as the
threshold increases up to 1.5$Q_{s}$.

For the dynamic threshold, a robust power-law behavior is observed.
The exponent $\theta$ of $\zeta=Q_{d}(t)$ is measured to be 0.58 for
the NYSE and 0.56 for the CSM respectively, somewhat close to
$\theta=1/2$ of the Gaussian white noises. But as $\zeta$ increase,
the network structure stabilizes, with the exponent $\theta=0.60$
for the NYSE and 0.61 for the CSM. The time-averaging degree
$\overline{K}$ is also found to decay slowly as the dynamic
threshold increases, and the network shows significant connections
for a high threshold as 6$Q_{d}(t)$, as shown in the inner panel of
figure \ref{Fig:5}(b) and (d).

Why is the dynamic threshold crucial in the analysis of the network
structure of financial markets? One important reason is that the
volatilities fluctuate strongly in the dynamic evolution. It induces
large temporal fluctuations of the cross-correlations of price
returns, as is characterized by $Q_{d}(t)$ in figure \ref{Fig:1}.
Thus the static threshold leads to dramatic changes in the
topological structure of the network. However, the dynamic threshold
proportional to $Q_{d}(t)$ suppresses such a kind of fluctuations,
and results in a stable topological structure of the network. For
example, we have calculated the time-averaging degree in two typical
periods of time for the CSM, i.e., $t = 1550$ to $1650$ with small
volatilities and $t=2530$ to $2630$ with large volatilities. For the
static threshold $\zeta=Q_{s}$, the time-averaging degree is $39$
and $98$ respectively, far away from the time-averaging degree
$\overline K=68$ in the total time interval. For the dynamic
threshold $\zeta=Q_{d}(t)$, however, it is $107$ and $104$
respectively, both around $\overline K=107$. Similar results are
obtained for the NYSE, and also for the clustering coefficient
$C(t)$.

\subsection{Degree distribution}

\begin{figure}[htb]
\centering
\includegraphics[width=8.5cm]{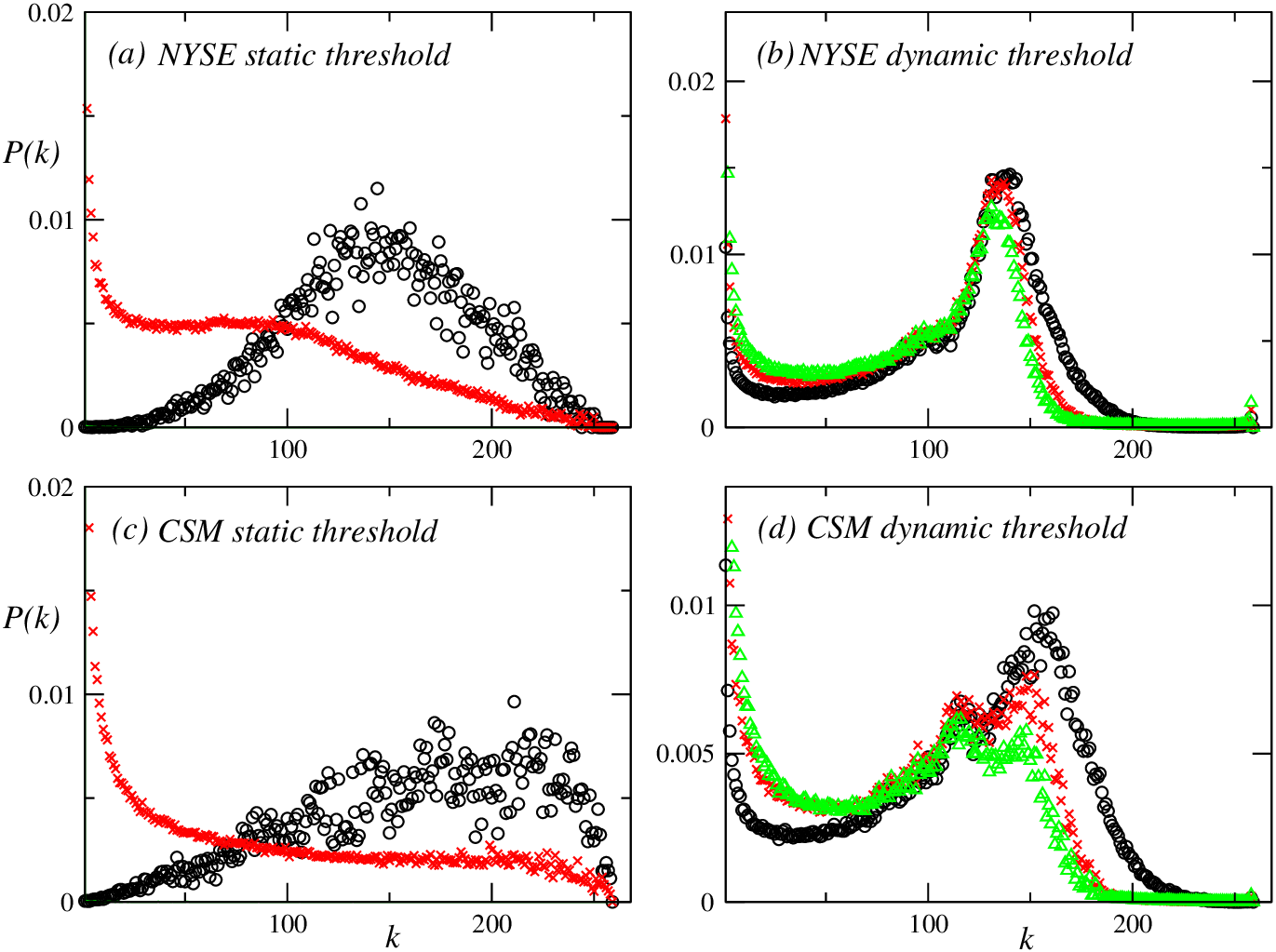}
\caption{\label{Fig:6} The degree distribution $P(k)$ is displayed
on a log-log scale. (a) and (c) are of the static thresholds for the
NYSE and the CSM. Circles and crosses are for $\zeta=0.5Q{_s}$ and
$Q{_s}$ respectively. (b) and (d) are of the dynamic thresholds for
the NYSE and the CSM. Circles, crosses and triangles are for
$\zeta=Q{_d}(t)$, 2$Q{_d}(t)$ and 3$Q_{d}(t)$, respectively.}
\end{figure}

The degree distribution function $P(k)$ describes the heterogeneous
properties of nodes. To obtain a better statistics, we take the
degrees of all time steps as an ensemble. Figure \ref{Fig:6} shows
the degree distribution of the static and dynamic thresholds for the
NYSE and the CSM. For both thresholds, it is observed that the
degree distribution is different not only from the Poisson
distribution, but also from the power-law distribution of a
scale-free network. For the static threshold, the degree
distribution is rather sensitive to the threshold value, and the
form of the degree distribution changes dramatically. For the
dynamic threshold, a two-peak behavior is observed with one peak
around $k=0$ and the other peak around $k=140 \sim 150$ for both the
NYSE and CSM. The peak at $k=0$ simply tells that there are a large
number of isolated nodes in the network. The peak at $k=140 \sim
150$ indicates that one node is directly connected to about one half
of the nodes in the network, i.e., the cross-correlation
distribution is hight-low symmetric for the dynamic threshold.

The two-peak structure of the degree distribution for the dynamic
threshold explains the upper envelope of the average degree $K(t)$
in figure \ref{Fig:4}. Moreover, the features of the degree
distribution is rather robust for different dynamic threshold
values. It indicates that the large cross-correlations are much
bigger than the average cross-correlation, so that certain increase
of $\zeta$ does not alter the network topology.

\subsection{Cross-correlation of degrees}

The so-called assortative or disassortative mixing on the degrees
refers to the cross-correlation of degrees \cite{new02,new03}. The
"assortative mixing" means that high-degree nodes tend to directly
connect with high-degree nodes, while the "disassortative mixing"
indicates that high-degree nodes prefer to directly connect with
low-degree nodes. The cross-correlation of degrees is defined as
\begin{equation}
r(t)=\frac{M^{-1}\sum_{\alpha}{j_{\alpha}k_{\alpha}}-
[M^{-1}\sum_{\alpha}{\frac{1}{2}(j_{\alpha}+k_{\alpha})}]^{2}}
{M^{-1}\sum_{\alpha}{\frac{1}{2}(j_{\alpha}^{2}+k_{\alpha}^{2})}-
[M^{-1}\sum_{\alpha}{\frac{1}{2}(j_{\alpha}+k_{\alpha})}]^{2}},
\label{e210}
\end{equation}
where $j_{\alpha}$, $k_{\alpha}$ are the degrees of the nodes at
both ends of the $\alpha_{th}$ edge, with $\alpha$=1,...,$M$. At a
certain time, $r>0$, $r=0$ and $r<0$ represent the assortative
mixing, no assortative mixing and disassortative mixing,
respectively.

\begin{figure}[htb]
\centering
\includegraphics[width=8.5cm]{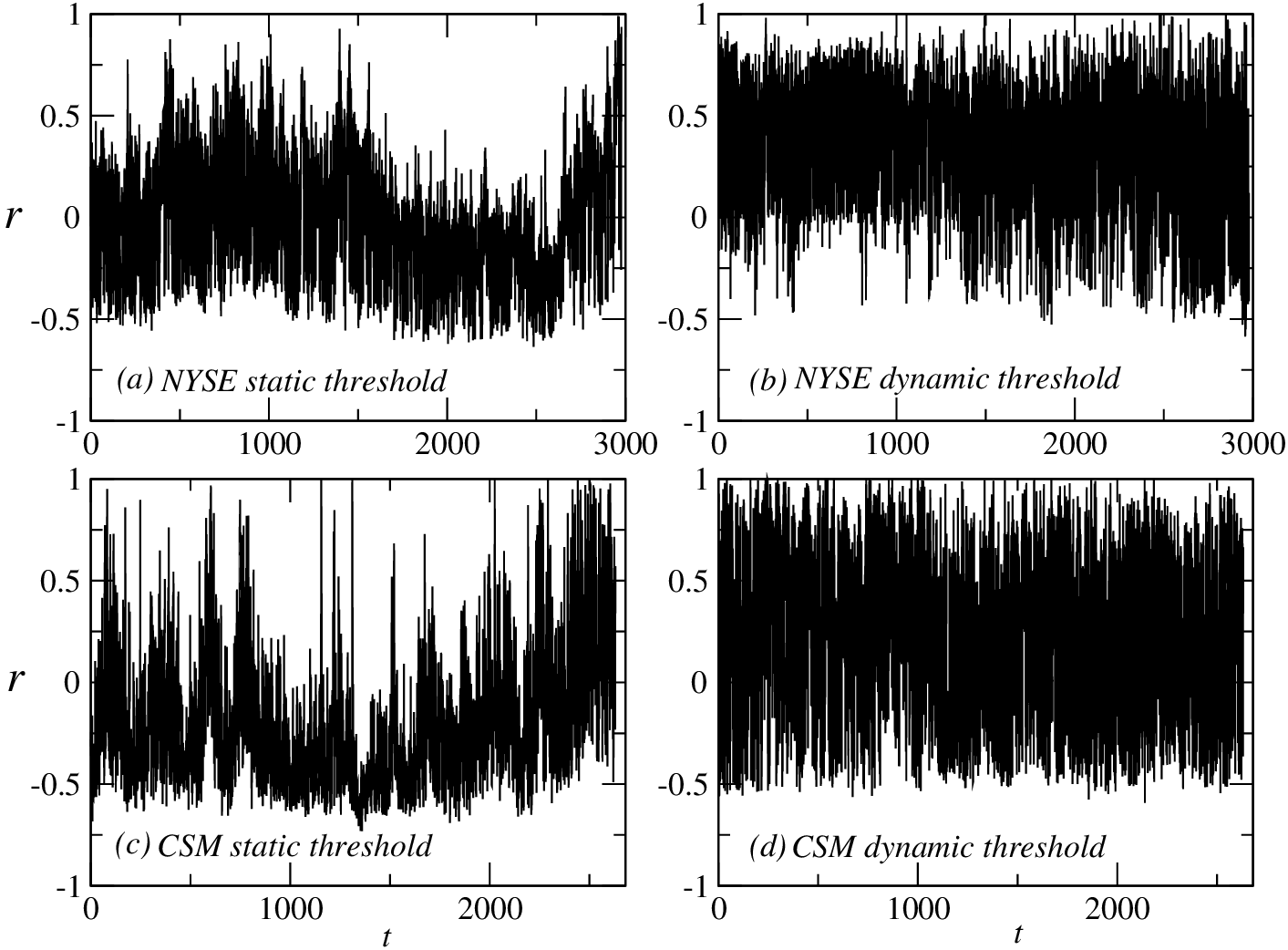}
\caption{\label{Fig:7} The cross-correlation $r(t)$ of degrees is
displayed. (a) and (c) are of the static threshold $\zeta=Q_{s}$ for
the NYSE and the CSM. (b) and (d) are of the dynamic threshold
$\zeta=Q_{d}(t)$ for the NYSE and the CSM. The time-averaging values
of $r(t)$ are $0.00$, $0.36$, $-0.20$ and $0.22$ respectively in
(a), (b), (c) and (d).}
\end{figure}

In figure \ref{Fig:7}, $r(t)$ of $\zeta=Q_{s}$ and $Q_{d}(t)$ is
shown for the NYSE and the CSM. $r(t)$ fluctuates between the
interval $[-1,1]$, flipping between the assortative mixing and
disassortative mixing during the time evolution. Defining the
time-averaging cross-correlation of the degrees,
\begin{equation}
\overline{r}=\frac{1}{T} \sum _{t=1}^{T} r(t),
\end{equation}
we obtain $\overline{r}=0.00$ and $-0.20$ of the static threshold
for the NYSE and the CSM, $\overline{r}=0.36$ and $0.22$ of the
dynamic threshold for the NYSE and the CSM. In other words, the
cross-correlation of the static threshold shows no assortative
mixing or the disassortative mixing, while that of the dynamic
threshold exhibits the assortative mixing.

\begin{figure}[htb]
\centering
\includegraphics[width=8.5cm]{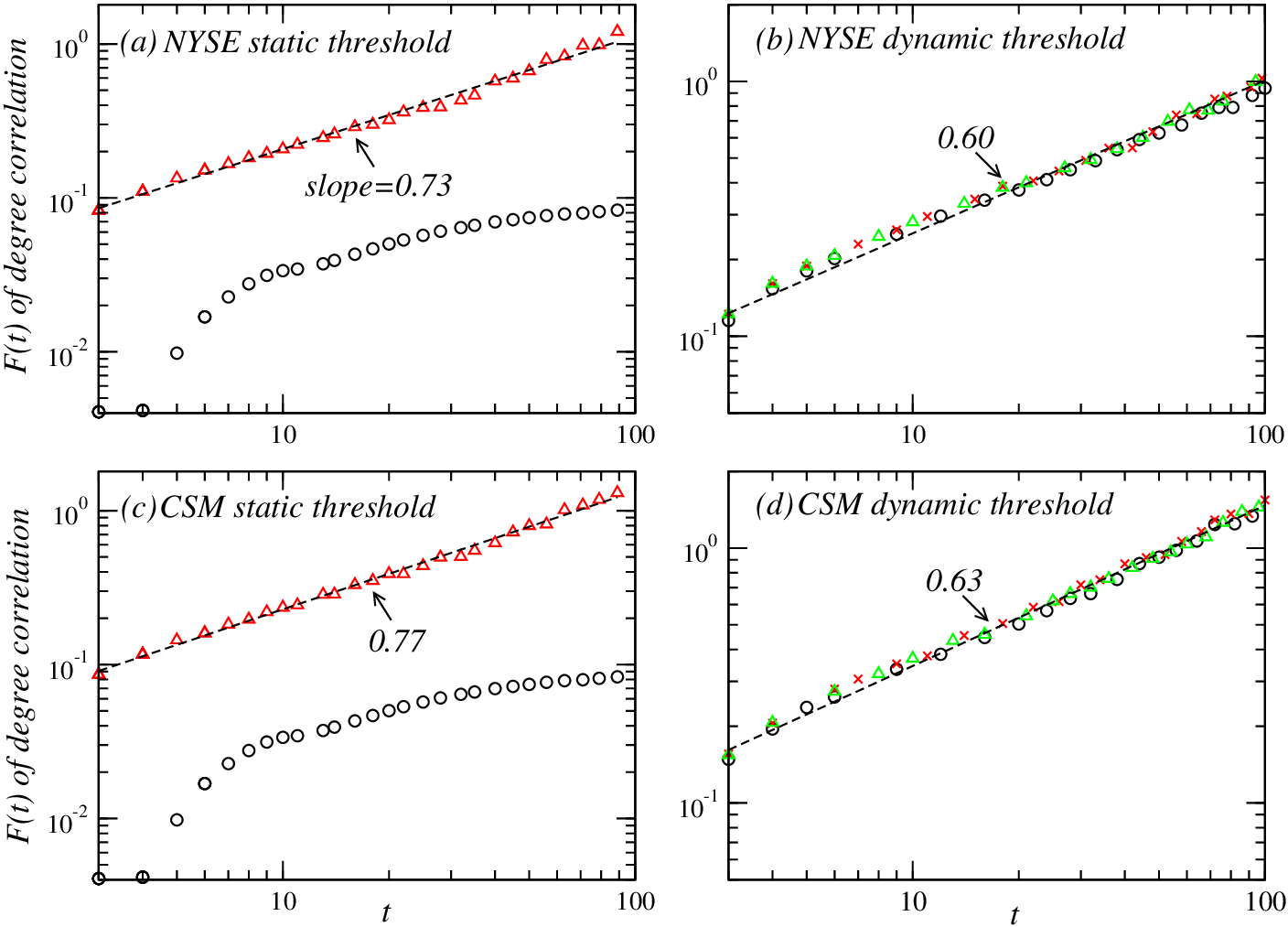}
\caption{\label{Fig:8} The DFA function of the cross-correlation of
degrees is plotted on a log-log scale. Dashed lines are the
power-law fits. (a) and (c) are of the static thresholds for the
NYSE and the CSM. Circles and triangles are for $\zeta=0.5Q_{s}$ and
$Q_{s}$ respectively. (b) and (d) are of the dynamic thresholds for
the NYSE and the CSM. Circles, crosses and triangles are for
$\zeta=Q{_d}(t)$, 2$Q{_d}(t)$ and 3$Q_{d}(t)$, respectively.}
\end{figure}

To study the memory effect, we again compute the DFA function of
$r(t)$. As shown in figure \ref{Fig:8}, the DFA function of
$\zeta=Q_{s}$ shows a power-law behavior with the exponent
$\theta=0.73$ for the NYSE and 0.77 for the CSM. For
$\zeta=0.5Q_{s}$, no power-law behavior is observed. It further
confirms the unstable network structure for the static threshold.
For the dynamic threshold, the DFA function of $r(t)$ shows nearly a
same power-law behavior for $\zeta=Q_{d}(t)$, 2$Q_{d}(t)$ and
3$Q_{d}(t)$, with the exponent $\theta=0.60$ for the NYSE and 0.63
for the CSM.

\begin{figure}[htb]
\centering
\includegraphics[width=8.5cm]{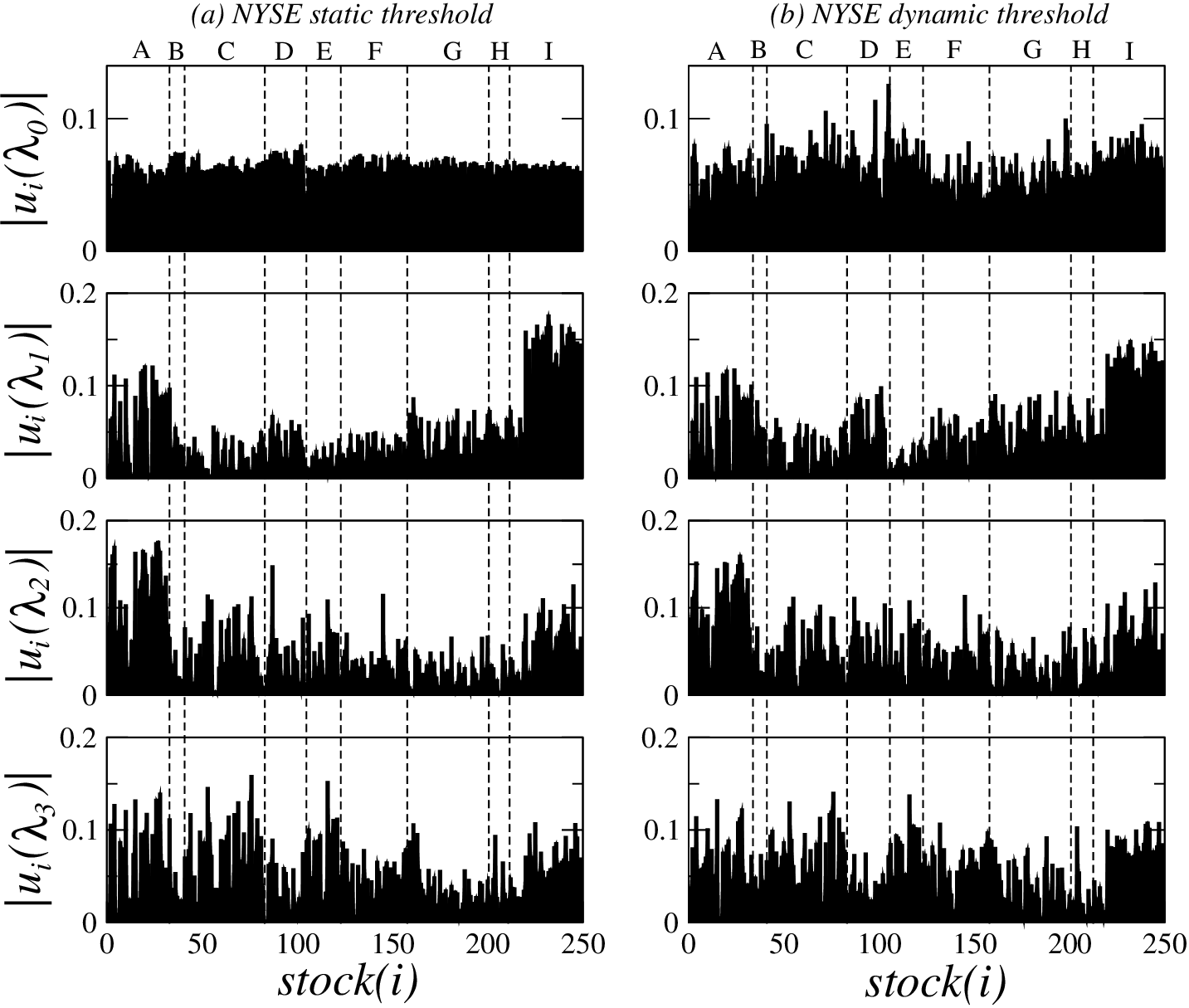}
\caption{\label{Fig:9} The absolute values of the components $u_{i}$
of stock $i$ for the first four largest eigenvalues of the
cross-correlation matrix $\mathbf{F}$ are displayed for the NYSE.
Stocks are arranged according to economic sectors separated by
dashed lines. A: Basic Materials; B: Conglomerates; C: Consumer
Goods; D: Finance; E: Healthcare; F: Industrial Goods; G: Services;
H: Technology; I: Utilities. (a) is of the static threshold
$\zeta=Q_{s}$, and (b) is of the dynamic threshold
$\zeta=Q{_d}(t)$.}
\end{figure}

The community or sector structure identifies different
interconnected subsets of networks
\cite{ple02,she09,cor05,pal05,pal07}. To further understand the
topological structure of the financial networks, we may investigate
the dynamic effect of economic sectors. The random matrix theory is
a representative approach to such a problem, e.g., one may analyze
the eigenvalues and eigenvectors of the cross-correlation matrix of
price returns \cite{ple02,she09,cor05}. With a similar procedure, we
first introduce the normalized individual degrees
$\widetilde{k}_{i}(t)= (k_{i}-<k_{i}>)/\sigma_{k_i}$, with
$\sigma_{k_i}$ being the standard deviation of $k_i$. We then
construct the cross-correlation matrix $\mathbf{F}$ of individual
degrees $\widetilde k_i(t)$, whose elements
\begin{equation}
F_{ij}=\frac{1}{T}\sum^{T}_{t=1} \widetilde{k}_{i}(t)
\widetilde{k}_{j}(t), \label{e310}
\end{equation}
and compute its eigenvalues and eigenvectors. For the network of the
NYSE, we do observe that the largest eigenvalue $\lambda_0$
corresponds to the market mode, while other large eigenvalues
$\lambda_i$ typically represent different economic sectors, as shown
in figure \ref{Fig:9}. For the dynamic threshold, the results are
robust when $\zeta$ changes from $Q_d(t)$ to $6Q_d(t)$. For the
static threshold, however, reasonable results are obtained only
around $\zeta=Q_s$. For the CSM, the dynamic effect of the standard
economic sectors is weak, and we need a careful analysis to reveal
its unusual sectors as reported in Refs. \cite{she09}.

\section{Conclusion}

We investigate the topology dynamics of a financial network by a
comparative study with static and dynamic thresholds, based on the
daily data of the American and Chinese stock markets. For both stock
markets, the dynamic threshold properly suppresses the large
fluctuation induced by the cross-correlations of individual stock
prices, and creates a rather robust and stable network structure
during the dynamic evolution, in comparison to the static threshold.
Long-range time-correlations are revealed for the average clustering
coefficient, the average degree and the cross-correlation of
degrees.

The average clustering coefficient and average degree for both
static and dynamic thresholds are large, indicating the strong
interactions between stocks in financial markets. A two-peak
behavior is observed in the degree distribution for the dynamic
threshold, very different from the power-law behavior of a
scale-free network.

\ack This work was supported in part by the National Natural Science
Foundation of China (Grant Nos. 10805025, 10875102, 10774080 and
10775071) and Zhejiang Provincial Natural Science Foundation of
China under Grant No. Z6090130.

\end{document}